# Towards a classification of Lindenmayer systems

Diego Gabriel Krivochen, University of Reading, CINN[1]

Douglas Saddy, University of Reading, CINN

Abstract:

In this paper we will attempt to classify Lindenmayer systems based on properties of sets of rules and the kind of strings those rules generate. This classification will be referred to as a 'parametrization' of the L-space: the L-space is the phase space in which all possible L-developments are represented. This space is infinite, because there is no halting algorithm for L-grammars; but it is also subjected to hard conditions, because there are grammars and developments which are not possible states of an L-system: a very well-known example is the space of *normal grammars*. Just as the space of normal grammars is parametrized into Regular, Context-Free, Context-Sensitive, and Unrestricted (with proper containment relations holding among them; see Chomsky, 1959: Theorem 1), we contend here that the L-space is a very rich landscape of grammars which cluster into kinds that are *not* mutually translatable.

Keywords: Normal grammars; Lindenmayer grammars; derivations; formal language theory

1. Introduction:

Let us first define a 'language', abstractly, as a mathematical system. This system has at least:

- A finite alphabet
- A set of states
- A transition function between states

In this view,

> *A sentence* over an alphabet is any string of finite length composed of symbols from the alphabet. Synonyms for sentence are *string* and *word.* (Hopcroft & Ullman, 1969: 1)

The grammar for a language *L*, call it *G(L)* is the set of rules over an alphabet which generates strings. More technically, grammars are *sets*:

$G(L) = (V_N, V_T, P, S)$, where

> *The symbols $V_N$, $V_T$, P, and S are, respectively, the variables, terminals, productions, and start symbol. $V_N$, $V_T$, and P are finite sets*. (Hopcroft and Ullman, 1969: 10)

The usual state of affairs in formal and natural language theories is to capture formal relations by means of grammars in Chomsky Normal Form (CNF):

> Chomsky-normal grammar: every context-free language is generated by a grammar for which all productions are of the form A → BC or A → *b*. (A, B, C, nonterminals, *b* a terminal)[2]

---

[1] Corresponding author: diegokrivochen@hotmail.com
[2] Chomsky-normality is equivalent to Greibach-normality, which defines all rules as being of the form A → *a*B or A → ε, where B is a variable over strings, *a* is a non-null terminal, and ε is the null symbol. See Greibach (1965).

This means that our alphabet needs two kinds of symbols: **non-terminals** (or 'intermediate symbols'), which get rewritten, and **terminals**, which do not get rewritten. A string (or a 'sentence', or a 'word') is a *linear concatenation of terminal symbols*.

In the present paper we will deal with a special kind of rewriting rule system: L-grammars. These systems are also of the form $\Sigma \rightarrow F$, like Phrase Structure Grammars of the kind described in Chomsky (1959) and Greibach (1965), but they have some special properties, both of which will be reviewed below:

a) Terminals / nonterminals are defined *contextually* within a rule, not in the alphabet
b) All possible rules involving elements in a representation apply simultaneously (i.e., there is no 'Traffic Convention')

Property (a) means that terminal and nonterminal nodes are not defined in the alphabet: in traditional computational terms, every member of the alphabet is at the same time a member of the set of initial states and of the set of accepting states. That is nearly inconceivable from a classical computational point of view, but the biological origin of L-grammars *qua* formalisms justifies this departure from classical assumptions. Property (b), on the other hand, implies a deviation from sequential rule systems, since everything that *can* be rewritten at each generation *is* effectively rewritten at the same time (Vitányi, 1978: 10). Moreover, the transition function in L-grammars imply that the cumulative nature of the derivation does *not* rest on subsequent generations existing at the same time (as opposed to Phrase Markers in which, say, a verb phrase contains *both* a VP node and its daughter nodes V and NP). Because rules apply to multiplication in biological organisms -like bacteria reproduction- or plant growth, once a cell divides –say- the original cell *does not exist anymore*. This follows directly from (a) and (b), and is essential to bear these properties in mind, as well as their implications, when working with L-systems from a formal point of view. Crucially for models of labelling (i.e., identification of nonterminals), L-systems do not allow backtracking of any kind, and thus any attempt of labelling nodes via Minimal Search algorithms (e.g., Chomsky, 2015) cannot apply. More generally, as we will insist on below, the concept of *label* is not formulable under L-grammatical assumptions, given (a) above. Note that the fact that more than a single element is acted upon at any time (as we rewrite *any and all symbols that we can rewrite*) make L-systems incompatible with computational systems built around Von Neumann's architecture: unless sequentiality is forced upon the L-grammar (that is, unless we make L-systems comply with some form of Traffic Convention), simultaneity in rule application makes L-systems orthogonal to any system based on Von Neumann's architecture, which crucially includes Turing Machines.

Apart from those unique properties, L-systems share many of the aspects that characterize other formal grammars: they have an alphabet, a set of states, and a transition function from one state to another. The peculiarities of L-grammars pertain to *how* they operate with those elements, which they have in common with other formal systems.

Derivationally, it must be noted that the simultaneity in L-systems contrasts drastically with the sequentiality in normal grammars. While only one rule can apply at a time *per generation* in a Chomsky-normal grammar, even if there is more than one nonterminal that can be rewritten[3], L-grammars rewrite *all* possible symbols *per* generation with all rules that can apply doing so *at the same time*, yielding a completely different growth pattern. CNGs operate from left-to-right, being based on the model of Turing Machines: these have an infinite one-dimensional memory tape, in which symbols are printed, replaced, or erased by a read-write head following instructions specified as the production rules of the grammar.

---

[3] In this respect, it is useful to refer to Lees' (1976) analysis of the formal conditions over *immediate constituent* approaches to structural descriptions, which are the basis for generative grammars, both transformational and non-transformational (see Schmerling, 1983 for further discussion about *immediate constituent* grammars). He concludes that the essential condition for the formulation of the rules of a grammar '*is simply that no more than one abstract grammatical symbol of a string be expanded by a given rule at a time*' (Lees, 1976: 30)

In this paper we will define and characterize each type of L-system, and provide examples of each. It is important to note that the present classification has not been devised analytically, and to our knowledge, the analytic derivation of properties of L-systems has by and large not been tackled in their 'non-normal' specificity[4], for L-systems have been mostly assimilated to normal grammars even though some derivational differences have indeed been acknowledged (e.g., Prusinkiewicz and Lindenmayer, 2010: 3, ff.; Prusinkiewicz and Hanan, 1989: 5, ff.; also Rozenberg and Salomaa, 1980). For the time being, thus, the present proposal is a refined heuristics, which poses interesting questions to be addressed from an analytical perspective. We will present some properties of *both* derivations and representations in L-systems, because both are relevant for the analysis of these grammars and their comparison with CNGs.

The systems we will be considering have two basic elements:

- An alphabet $\Sigma$
- A set of transition rules over $\Sigma$

More formally, a Lindenmayer grammar G is a set $G = (R, \Sigma)$, where R is a set of transition rules and $\Sigma$ is the alphabet. Both of these are somewhat exceptional in the L-systems realm. To begin with, whereas the definition of normal grammars includes the specification of a set of initial states and a set of final (or accepting) states (Hopcroft and Ullman, 1969), in the case of L-grammars any symbol is *at the same time* an initial and an accepting state, meaning that any symbol can appear at the left *and* right-hand side of a transition rule. For example:

1) $1 \rightarrow 0\ 1$

In the context of an L-grammar, it is not possible to say '1 is a terminal' or '1 is a nonterminal', because those notions only make sense within normal grammars. For the purposes of emergent properties of L-grammars, *all symbols are considered terminals*, yet all symbols *can* rewrite (but we will see not all *actually do*). This means L-systems lack the representational stability that so-called *normal* grammars have, as *there is no labelling procedure* in L-systems: labels cannot exist, because of the simultaneous and obligatory rewriting of all symbols in any given generation. All these properties make L-systems orthogonal to the Chomsky Hierarchy of formal languages (Chomsky, 1950).

Then, the set of transition rules also differs from what is customary in normal grammars. Prusinkiewicz and Lindenmayer (2010) expose the issue neatly:

> *In Chomsky grammars productions are applied sequentially, whereas in L-systems they are applied in parallel and simultaneously replace all letters in a given word. This difference reflects the biological motivation of L-systems. Productions are intended to capture cell*

---

[4] There are, of course, some exceptions. A recent one is Patel et al. (2015).

> *divisions in multicellular organisms, where many divisions may occur at the same time.*
> (Prusinkiewicz and Lindenmayer, 2010: 3)

This has far-reaching consequences for computational modelling of L-systems: a cursor-based algorithm which 'counts' symbols in a string and applies a rule as soon as the first suitable input is found cannot yield the same results as an L-system, in which *all symbols rewrite simultaneously*. Such a simultaneous procedure is in essence different from an *immediate-constituent* (IC) analysis of strings, in which a condition for the formulation of the rules of a grammar '*is simply that no more than one abstract grammatical symbol of a string be expanded by a given rule at a time*' (Lees, 1976: 30). Equivalently, Greibach (1965: 43) defines one of the properties of a phrase structure derivation in the following terms:

> *All generations proceed from left to right, expanding the left-most member of I* [the set of nonterminal symbols] *first.*

Let us see an example of the sequentiality of Chomsky-normal grammars –all of which model IC processes-, taken from Chomsky himself (1957: 26):

2) (i) Sentence → NP + VP
   (ii) NP → T + N
   (iii) VP → Verb + NP
   (iv) T → the
   (v) N → man, ball, etc.
   (vi) Verb → hit, took, etc.

Developing this set of rules according to Chomsky's own derivation, we get:

3) Sentence (Axiom)
   NP + VP (by (i))
   T + N + VP (by (ii))
   T + N + Verb + NP (by (iii))
   The + N + Verb + NP (by (iv))
   The + man + Verb + NP (by (v))
   The + man + hit + NP (by (vi))
   The + man + hit + T + N (by (ii))
   The + man + hit + the + N (by (iv))
   The + man + hit + the + ball (by (v))

A single step may illustrate our point: a simultaneous rewrite system would give us [The + N + Verb + the + N] in step 4, but the system proceeds left-to-right, finds the first nonterminal that can be subjected to a rule of the set (i-vi), applies it, and the newly generated string is the input for the following derivational step. Moreover, since [hit], [the], etc. have no transition rule associated to them (i.e., they are terminal symbols), the computation halts once the string only contains accepting states. On the contrary, if the terminal / non-terminal character of a symbol is defined contextually within a rule, as in L-systems, then the procedure never needs to halt. A derivation like (3), thus, is impossible to model in L-grammatical terms, because it is dependent on the distinction between terminal and nonterminal nodes *in the lexicon* and the fact that terminal nodes never appear at the left of rewriting rules.

In what follows, we will make reference to the form of rules in specific L-grammars, more specifically, we will need to make generalizations about the *right-hand* side of rules. To this end, we need to introduce some notation:

Let an L-grammar $G$ be a set of rules $R = \{r_{1i}, r_{2j}, ..., r_{nk}\}$. Because the left-hand side of the rules will always contain a single symbol (we will see why below), let $r_n \in R$ be a proxy for the symbols in the right-hand side of $r_n$ with the index $i, j, k...$ denoting the number of *non-null* symbols in the right-hand side of $r$. Let us give an example. Consider the following grammar, given an alphabet $\Sigma = \{A, B\}$:

4) $A \rightarrow B\ A$
   $B \rightarrow A\ B$

This L-grammar (let us call it $G_1$) has two rules, $r_{1i}$ ($A \rightarrow B\ A$) and $r_{2j}$ ($B \rightarrow A\ B$). Since the right-hand side of both rules contains two elements, $i = j = 2$. The rules differ in the order of the symbols in their right-hand side, which is why we have two different $r$'s. In what follows, we will drop the numbering of the rules (because we will conceive of the grammar as an unordered set), and *only refer to their index*.

We only consider *non-null* members of the alphabet $\Sigma$. Thus, if we consider (5):

5) $A \rightarrow B\ \varepsilon$
   $B \rightarrow A\ B$

We have $r_i$ (B, $\varepsilon$) and $r_j$ (A, B) for $i = 1$ and $j = 2$, because $\varepsilon$ is the radically empty string.

In all cases, we will assume that the grammar does *not* halt after a finite number of steps: this implies that at least one member of $\Sigma$ appears in *both* the left and right-hand sides of the rules. This is obviously not required in normal grammars, as we saw in the derivation (3); in this case, lexical insertion marks the halting point: having only terminals is the natural way of terminating derivations in CNGs.

Since we have pointed out the main differences between normal grammars and L-grammars, thus delimiting the space of possible grammars and restricting it accordingly, let us now get into the classification without further ado:

2. <u>Symmetric</u> (e.g. XOR; XOR[†])

An L-grammar $G$ is said to be *symmetric* if $G = \{r_i, r_j, ..., r_n\}$ for $i = j = n$ **and** every member of $\Sigma$ rewrites as a non-null symbol in all $r$.

Both clauses are necessary conditions, recall that only non-null members of the alphabet count for our purposes.

The segmentation of the string corresponding to any generation $g_n$ can be expressed in terms of operations over strings corresponding to previous generations: this is possible because L-systems can be used to model recurrence relations. We will work with two such operations, which we will refer to as *mappings* from strings to strings:

$g^M$ = the *mirror* of generation $g$ (e.g., $(10110)^M = 01101$)

$g^N$ = the *negative* of generation $g$ (e.g., $(10110)^N = 01001$)

These are both *homomorphic* mappings: they preserve relations within selected structure. *M* and *N do not* affect the emergent properties of a grammar for any set of strings generated by that grammar in subsequent derivational steps.

This could mean that emergent properties are tied to growth properties of the grammar rather than to properties of the alphabet (e.g., number of symbols). It is crucial to note that the mappings that are relevant for generation *N* apply to *N-1* and *N*, and only to them. This is not surprising if we recall the nature of L-grammars: they are defined as *recurrence relations*[5]. Moreover, since the system is strongly derivational, and there is no representational stability that can give rise to labelling, it is impossible to hang on to a generation that has already been subject to rewriting: a crucial difference between L-systems and grammars in CNF is that *N-2* no longer exists at the point of applying the relevant rules at N. The mappings, thus, are relevant *within* the transition between *N-1* and *N*.

Now, what happens if we want to map a string onto a copy of that string? That is, for example,

6) 10110 | 10110

The string in (6) is divided in two equally long substrings (which we indicate by a vertical line | ), which do not display, *prima facie*, either M or N mappings. Rather, the substrings are identical copies, which are concatenated. Do we need to propose a third mapping, Copy(*g*), which yields another instance of *g*?

$$g^C = g$$

Optimally, we would try to avoid that, keeping the number of mappings to a minimum. But how do we establish which mappings are independently required and which should be expressed in terms of combinations of more basic mappings?

Computationally, it seems that Copy is simpler than M and N (since we do not modify the copied substring in any way, we just repeat it as it is), and probably it is developmentally more basic also (phylogenetically speaking); why haven't we distinguished Copy as a separate primitive operation? Because Copy can in principle be decomposed in terms of the recursive application of other operations, whereas M and N cannot. Copy can be expressed in terms of M(M) or N(N); thus, by redundancy rules applying to the meta-theory, Copy can cease to exist as an independent, first-order operation to be a complex, second-order operation -for it can be subsumed to the newer, more powerful M and N homomorphic mappings. Let us see an example:

7) $(10110)^M = 01101$ $\qquad (10110)^N = 01001$
   $(01101)^M = 10110$ $\qquad (01001)^N = 10110$

We can concatenate *g* with $g^M$, $g^N$, $(g^M)^N$, $(g^N)^M$, and, more generally, *g* affected by a potentially unlimited number of instances of the mappings. The mappings are themselves an interesting object of

---

[5] The *Encyclopedia of Mathematics* (http://www.encyclopediaofmath.org/index.php?title=Recurrence_relation&oldid=32959) defines a *recurrence relation* as

> *A relation of the form*
> $a_{n+p}=F(n, a_n,...,a_{n+p-1})$,
> *permitting one to compute all members of the sequence $a_1, a_2,...,$ if its first p members are given.*

study, if they are real conditions over the development of the grammar. They display the following conditions, tested by brute force:

8) *M* and *N* are *commutative* (i.e., for *x* a string derived via an asymmetric L-grammar, $(x^M)^N = (x^N)^M$)
   *M* and *N* are not monotonically recursive (i.e., $(x^N)^N = x$ and $(x^M)^M = x$ for all *x*): this property lets us recast Copy in terms of sequential applications of the same mapping to a given (sub)string.

Let us see an example, in which the homomorphic mappings are color-coded (grey for M, black for N). Each generation $g_n$ is formed by repeating the previous generation $g_{n-1}$ and right-adjoining to that either the mirror or the negative of $g_{n-1}$:

XOR (a.k.a. Thue-Morse grammar)

0 → 0 1
1 → 1 0

```
                                0
                               01
                              0110
                           01101001
                       0110100110010110
               01101001100101101001011001101001
01101001100101101001011001101001 1001011001101001011010011001 0110
```

XOR†

0 → 1 0
1 → 0 1

```
                                0
                               10
                              0110
                            10010110
                        0110100110010110
                   10010110011010010110100110010110
01101001100101101001011001101001 1001011001101001011010011001 0110
```

Notice, incidentally, how the N substrings in XOR and XOR† are the N of each other: the N in XOR-$g_3$ is itself the N of XOR†-$g_3$. This is precisely what we would expect if M and N indeed behave as homomorphic mappings, because the selected structure is preserved between the grammars. Some properties of Thue-Morse strings have been studied in similar terms to those proposed here: Brlek (1989: 84) introduces the notion of the *complement* Ū of a Thue-Morse word U as the result of exchanging the members of the alphabet in the string. This notion, which is equivalent to our *negative* for a binary alphabet, is useful in the analysis of recurrent substrings within the Thue-Morse word. In general, for a *binary* alphabet, there are infinite *squares* (i.e., juxtapositions of two identical substrings) in an infinite string; however, there are no *cubes* (juxtapositions of three identical substrings). Other relevant properties also hold at the limit, when the length of the string approaches infinity: the *critical exponent*[6] of the infinite Thue-Morse word is 2 (Thue, 1912; Krieger, 2007).

These sequences present the same number of '1' and '0' in each generation, which means that the ratio between members of Σ is *constant*. There is no bias in the distribution of elements of the alphabet in the grammars, although this does not mean the grammar has entropy = 0. As a matter of fact, a '1' can be followed by either '0' or '1', and a '0' can be followed by either '1' or '0'. However, any attempt to come up with a statistical method as a way to 'infer' what can come next would fail, because the XOR space is evenly distributed between '0' and '1'.

3. Asymmetric (e.g., Fib, Lucas, ε-Fib)

An L-grammar *G* is said to be asymmetric if:

i. Weak condition: $G = \{r_i, r_j, ...r_n\}$ for $0 < r_i < r_j < ...r_n$

ii. Strong condition: Same as i. with the added requirement that $r_i \subsetneq r_j \subsetneq ...r_n$.

We will refer to the rule whose right-hand side properly contains the right-hand side of another rule as a 'strong' term. So, for instance, if we consider the L-grammar $G_1$ with Σ = {0, 1} and rules:

$r_i = 0 \rightarrow 1\ 0\ (i = 2)$

$r_j = 1 \rightarrow 1\ 0\ 1\ (j = 3)$

We see that $r_2 \subsetneq r_3$ (recall that the subindexes refer to the number of non-null symbols in the right-hand side of the rule), as $r_3$ properly contains [1 0] plus an extra '1'. In this case, $r_3$ is our 'strong' term.

In the manipulation of L-systems (when expanding and pruning the trees that these grammars generate), we can delete freely anything that rewrites as ε and anything that does not refer to the left-hand side of the rules, which is why we initially said that we would ignore null symbols. Non-null symbols which do not rewrite make the grammar 'halt' locally, and impact on the space-filling properties of the system; these symbols will be referred to as 'stumps'. As is the case with CNGs, the quantitative properties of L-words *w* are clearer as $|w| \rightarrow \infty$ (a.k.a., 'at the limit'. See also Matlach and Krivochen, 2016 for a quantitative analysis of some properties of formal and natural language strings).

The conditions ensure that there is a 'weak' term and a 'strong' term; in other words, the space defined by the grammar is *not* evenly distributed among the elements of the alphabet as $|w| \rightarrow \infty$. There are conditions over what the strong term can contain, which introduce further distinctions within the space of asymmetric L-systems. The strong term properly contains the weak term in the following sense: whatever the weak symbol rewrites as, it appears as part of what the strong symbol rewrites as, *plus* 'something else'. It is the nature of this 'remainder' that determines the emergent properties of the relevant L-grammar. Let us see an example:

9) $a \rightarrow [a\ b]$
$b \rightarrow [a\ b]\ b\ a\ b$

---

[6] The critical exponent of a sequence describes the largest number of times a contiguous subsequence can be repeated. Relevantly, it need not be an integer or even a rational number, as we will see below.

In (9), *a* is the weak symbol, and *b* is the strong symbol. We see that *b* rewrites as a string that *properly includes* whatever *a* rewrites as (in this case, [*ab*]) plus the substring *bab*: this substring is what we called the 'remainder'. We will come back to the relation between strong terms, weak terms, and remainders shortly.

An interesting property to be highlighted of this kind of L-systems is that *in the derivation of an asymmetric L-grammar satisfying the strong condition, at each generation the number of occurrences of each non-null symbol is a member of the Fibonacci sequence* ('Fib' henceforth) under specific conditions over the remainder which we will specify below. Fib arises as well as a 'syntactic' result (Uriagereka, 2015) of considering the number of elements per generation, without discriminating among them. In the following Fib trees, we will consider the number of '1' per generation, but the reader is welcomed to count just the '0's, as well as the total number of symbols:

Fib-grammar

$0 \to 1$
$1 \to 0\ 1$

```
                          0
                          1                          1
                         01                          1
                        101                          2
                       01101                         3
                      10101101                       5
                    0110110101101                    8
                 101011010110110101101                13
             0110110101101101011010110110101101      21
```

Trivially, the Fib-grammar can be 'expanded' as:

$0 \to 1\ \varnothing$
$1 \to 0\ 1\ \varnothing$

In which Ø, naturally, does not rewrite. This is a purely formal point which might be ignored for the purposes of biological implementations of L-grammars, but it turns out to be relevant for the emergent properties of L-systems *qua* formal grammars. Note that this expansion maintains the condition that the strong term properly contains the weak term (plus the axiom). The symbols that do not rewrite are *constants*, and by virtue of not developing into anything else, they represent points at which the system *locally halts*. In a view in which the growth of an L-grammar fills a space, these symbols (which we call *stumps*, a term that is due to Juan Uriagereka) create local empty spaces, points (or sets thereof) that the grammar does *not* fill. The density with which these stumps occur heavily impacts on the space-filling properties of a given L-grammar at the limit.

The simplest possible Fib-grammar (which we will refer to as the *minimal Fib grammar*) has the following properties:

- It has a non-branching term (the 'weak' term) and a branching term (the 'strong' term)
- The non-branching term contains the 'seed' of the branching term (i.e., its left-hand side)
- The branching term contains all non-null members of Σ (including, of course, the weak term)

We need to be able to reduce any grammar that is more complex (in terms of involving more rules and/or elements in the alphabet) to this minimal expression if we want it to generate Fib. It seems that, when expanding the minimal Fib-grammar –and if we want to maintain its emergent properties-, we can:

  a) Add and eliminate constants freely (e.g., the ε-Fib expansions below)

  b) Consistently permute 1s and 0s in the rules

  c) Make the right-hand side of the strong term (the branching term in the minimal Fib grammar) into a term itself (so that [0 1] → [1 0 1])

  d) Expand the grammar assuming some sort of constituency (or, rather, the atomization of whatever the axiom immediately dominates, which makes sense if it is the axiom of the grammar). This kind of expansion mechanism is particularly interesting, as it implies assuming some representational stability: we need to keep a symbol or a string 'in mind' (or in a memory stack) to give us the transitive expansion. For example, assume that we expand the irreducible Fib grammar such that the axiom 0 rewrites as something it dominates transitively, like generation 3 = [1 0 1]; then, the strong term rewrites following the rules above (containing all non-null members of the alphabet):

10)  0 → 1 0 1
     1 → [1 0 1] 0 1

It is time now to specify the relation between weak and strong terms in the rewriting system: whatever the weak term rewrites as, the strong term must contain it, plus a remainder which *has* to be a Fib-constituent for the sequence to arise[7]. So, predictably,

11) 0 → 0 1
    1 → [0 1] 0 1 0

Fails to deliver the Fib sequence and it also presents the Fib-illegal bi-gram [0 0]. The tweak in (11) results in a Fib-ungrammatical derivation because [0 1 0] is *not* a Fib constituent ([0 1] is, though, and so is [1 0] if we apply the mapping M or N; but under no matrix transformation can [0 1 0] arise as a natural constituent of the Fib grammar). This means that the irreducible Fib grammar sort of *folds into itself*, displaying crucial properties of structure preservation in a sense reminiscent to Emonds (1970): as long as we tamper with the right-hand side of the rules using substrings from the grammar which arise as 'constituents' in Fib, the emergent properties of the irreducible grammar will hold.

A further note is necessary here: recall that L-systems are essentially recurrence relations, which means that once the initial state is given, the state of the system at any point is defined as a function of the preceding states. In the case at hand, $Fib_n = Fib_{n-1} + Fib_{n-2}$. This gives us two kinds of Fib expansions of type (d):

- We make 0 rewrite as $Fib_n$ and 1 rewrite as $Fib_{n+1}$
- We make 0 rewrite as $Fib_n$ and 1 rewrite as $Fib_m$, where $m \neq n+1$ and $m \neq n-1$

For instance:

---

[7] This condition can be expressed as a constraint over auxiliary trees in a TAG-like L-system, see below.

12) a. 0 → 101 (= $Fib_3$)
    1 → 01101 (= $Fib_4$ = $Fib_2 \frown Fib_3$)
  b. 0 → 101 (= $Fib_3$)
    1 → 10101101 (= $Fib_5$ = $Fib_3 \frown Fib_4$ = $Fib_3 \frown Fib_2 \frown Fib_3$)

Both grammars generate Fib-grammatical strings (i.e., the illegal *n*-grams *00 and *111 are never generated), but the distribution of legal *n*-grams varies slightly, as does the growth function of the tree. We will refer to the expansion that use immediately subsequent generations (like (12 a), in which '0' rewrites as generation 3 of the irreducible Fib grammar and '1' rewrites as the immediately next generation) as a *Non-Skip* expansion, and to the expansion that uses non-subsequent generations (jumping over one or more in the rewriting, see (12 b), in which '0' rewrites as generation 3 of the irreducible Fib grammar, but '1' rewrites as generation 5, thus 'jumping over' generation 4) as a *Skip* expansion. Only the grammar that has been expanded from Fib by using subsequent generations as right-hand sides maps perfectly to Fib strings. This is particularly relevant if L-grammars are used to generate stimuli for experimental purposes (see the seminal work of Saddy, 2009 and much subsequent work).

Some operations over trees have been proposed by Uriagereka (2015) and Krivochen (2017) in an attempt to reduce expressions of more complex grammars to the fundamental and irreducible Fib-system. If such a reduction was possible, these authors conjecture, then the grammars are strongly equivalent to the minimal Fib grammar. The operations they proposed are the following:

Uriagereka's (2015: 668)

> Pruning: A non-branching symbol can be ignored in certain designated contexts [adjacent to or immediately dominating an atomized '1']

> Atomization: Any string of sister symbols [other than those involved in *pruning*] can be atomized into a single constituent symbol

Krivochen's (2017: 70)

> Collapse: when 0 is immediately dominated by a branching 0, collapse the former with its *non-empty* sister (thus a putative pair (0, ε), where ε is the empty symbol in Σ, does not collapse)

> Percolate: rewrite a branching 0 immediately dominating (0, *x*) as *x* iff *x* is *non-empty* (once again, (0, ε) does not get rewritten)

One of the problems, perhaps the simplest to express without actually drawing the graph, is choosing the directionality of the application of *collapse* and *percolate* (or *atomization* and *pruning*). Do we start from a rule (in which we can see what branches and what does not) and tamper with that rule? Or do we apply the rule as it is and then tamper with the resulting branching structure? In either case, at which generation do we start applying these 'transformations'? If we look at a rule, we could start the derivation with the modified grammar…but from a real-time perspective that would require that we should take into consideration a pair of objects (the objects on the right side of the rule, the 'F' part in a Σ, F grammar) which, at the point of applying the rule to the axiom, *do not yet exist*. Since L-systems are crucially *time-dependent*, the choice regarding what to operate on has far-reaching consequences: we could even talk about hypersensitivity to initial conditions, insofar as a single transformation operating on a single branching node could greatly alter the emergent properties of the L-system. This is related to a more fundamental property of (mapping) operations: they always apply to *representations*. This should come as no surprise, since processes cannot apply to other processes (or, in other words, a function cannot take a function as its only argument), but must apply to an object or set thereof. Summarizing: neither *collapse* nor *percolate* can apply *derivationally*. And this is profoundly 'counter L-intuitive', because, unlike normal grammars, L-systems are *strictly* derivational: as there are no terminal / nonterminal distinction, there is no 'representational stability', the system cannot get hold of an object in the form of a label, because *everything* is to be rewritten. As a result, strictly speaking, previous generations *do no longer exist*, and thus they (or any of its members) cannot be tampered with. We can consider the possibility of reduction (general 'tree pruning', in Ross' 1969 terms) via Uriagereka's or Krivochen's operations, as long as we bear in mind that we are imposing normality over a non-normal, simultaneous rewriting system.

It is always possible to expand an L-grammar using the methods explained above and then prune it by removing the added elements, which if we want to maintain the emergent properties of the grammar need to be subsequent Fib-generations. In general, if given a certain rule we can prune its right-hand side by constituent chunks (i.e., whatever we prune has to be a natural Fib constituent, that is, any of its generations) and we get a rule of the irreducible Fib grammar, we say that the pruned rule is equivalent to the corresponding Fib rule. The following two grammars are, in this sense, equivalent:

13) $G_1$:
   $0 \rightarrow 1\ [1\ 0\ 1]$
   $1 \rightarrow [1\ 0\ 1]\ 0\ 1$

   $G_2$:
   $0 \rightarrow 1$ (by deleting the Fib-constituent [1 0 1])
   $1 \rightarrow 0\ 1$ (id. ant.)

Equally, we can concatenate Fib generations to come up with new, expanded rules. The strings generated by this new grammar will be Fib-grammatical, and will indeed have numbers of 1s and 0s (and total symbols) per generation which belong to the Fib sequence:

14) $0 \rightarrow [0\ 1\ 1\ 0\ 1]\ [1\ 0\ 1]$  (= $g_3 + g_2$ in Fib)
    $1 \rightarrow [1\ 0\ 1\ 0\ 1\ 1\ 0\ 1]\ [0\ 1\ 1\ 0\ 1]$ (= $g_4 + g_3$ in Fib)

So far, we have been analysing what happens if we tweak the *right-hand* side of the rules. If we tamper with the *left-hand* side of the rules, replacing '0' or '1' with longer strings (e.g., replace {1} with {0 1}), we are imposing a non-natural notion of constituency, and grouping a substring of symbols for rewriting purposes: this is equivalent to *labelling* a substring (i.e., we need to establish that {0, 1} is actually <1, 0>, with an order imposed over the set: <1, 0> = <1, <1, 0>> by the *pairing axiom* in Zermelo-Fraenkel set theory; just like <V, N> = <V, <V, N>> in natural language phrase markers, see Fortuny, 2009). This operation amounts to artificially normalizing what is by nature not normal, and the result is a *dynamically frustrated* system. Let us develop this point with some more detail. Suppose we have a grammar that includes the rules 010 → 10110 and 011 → 01001. That is equivalent to a context-sensitive system, in which 1 followed by and preceded by 0 rewrites as 10110 and 1 followed by 1 and preceded by 0 rewrites as 01001. Now we take a look at the state of affairs in CNGs. The general format of a CS-rule is X → Y / W_Z, or, alternatively, WXZ → WYZ (Prusinkiewicz & Lindenmayer, 2006: 30 use the notation $a_l < a > a_r \rightarrow \chi$ to indicate that *a* rewrites as χ in context $a_l\_a_r$, thus we have $a_l a a_r \rightarrow a_l \chi a_r$). In either case, the read-write head must consider *more than a single symbol* when selecting the new symbol to be printed. Strictly speaking, and given enough memory, contexts can be complex phrases or words of any length up to the memory limits (Uriagereka, 2008: Chapter 6 presents a reinterpretation of the Chomsky Hierarchy in terms of memory capacities, terms in which is useful to think about this problem). In any kind of automaton based on Von Neumann's architecture, this presents no problem: the tape is read from left to right and the read-write head acts as a cursor. Thus, ambiguities never arise. But this is *not* the case for L-systems, where the left-hand side of rewriting rules must always consist of a single symbol to ensure the system will keep going on its own, without the need to resort to an external controller to decide what to do given an ambiguous configuration. The reason is that there is more than a single possibility to define and identify constituency for local *n*-grams, and thus what gets rewritten can be more than one substring at a time: note that this is a problem in a rewriting system that is defined by the *simultaneous* application of all possible rules (and is thus 'top-bottom' rather than 'left-to-right', unlike Chomsky/Greibach-Normal grammars). Consider, for instance:

15) 0 1 → 1 0 1
    1 0 → 0 1 0 1

If we are presented with a substring generated by the grammar in (15), a deterministic parser cannot automatically decide what to rewrite, particularly because L-systems do not present the left-to-right directionality in rule application (a.k.a., the Traffic Convention) that CNGs do (cf. the development in (3)): derivations in L-systems are strictly 'top-down', without any lateral orientation (and this is what allows all rules to apply simultaneously instead of sequentially). In L-systems there is no inherent hierarchy between the rules, that is, there is nothing that determines that rule X should apply *before* or *after* rule Y, contrarily to what happens in grammars in CNF: we cannot apply lexical insertion of [ball] to the terminal N if we haven't first rewritten NP as Det + N, for instance. In a system of the kind we analyze here, under the conditions in (15), every 2-*gram* generates a conflict between both rules, which can –and thus, *must*- apply to the *same object at the same time*, as we have seen above. Moreover, there is no way of biasing the derivational process once it has started (because we are dealing with a computable function implemented through an *a*-machine, there is no external controller that can arbitrarily decide which rule should apply), consequently, what we get is effectively a *locally and dynamically frustrated system*[8]:

16) …or this?
    0 1 0 1 0 1 0 1 0 1…
    Rewrite this…

The frustration illustrated in substring (16) of course extends all throughout the string; this should serve as an argument for our claim that the left-hand side of L-rules cannot be tampered with if we aim to construct a deterministic L-system that does not require an external controller. Let us now see an example of an asymmetric L-grammar that *only* satisfies the *weak* condition in the definition above:

ε-Fib 1

0 → 1 ε

---

[8] The situation in (16) illustrates a kind of *computational frustration* that was not taken into consideration by Binder (2008), who focused on DTMs (that is, on Chomsky Normal Grammars and their associated automata). What we can show is that a computational frustration can arise under simpler conditions, and even outside the space defined by normal grammars.

1 → 0 1

```
                                  0
                                 1ε                        1
                                 01                        1
                                1ε01                       2
                               011ε01                      3
                             1ε01011ε01                    5
                          011ε011ε01011ε01                 8
                      1ε01011ε01011ε011ε01011ε01          13
                 011ε011ε01011ε011ε01011ε01011ε011ε01011ε01   21
```

Note that, even though *i = j*, the symbol ε does not rewrite. In this case, ε is what we have referred to as a 'stump'. Interestingly, the emergent property of the Fib grammar is maintained, namely, the number of 0, 1, and ε *per generation* are members of the Fibonacci sequence (we have only indicated the number of 1s per generation, but the reader is welcomed to count 0 and ε as well). This is compatible with our characterization of asymmetric L-systems.

If we modify the rules so that the grammar satisfies *both* conditions, Fib still arises (and, again, we are just indicating the number of 1 per generation):

ε-Fib 2

0 → 1 ε
1 → 0 1 ε

```
                                  0
                                 1ε                        1
                                 01ε                       1
                                1ε01ε                      2
                               01ε1ε01ε                    3
                             1ε01ε01ε1ε01ε                 5
                          01ε1ε01ε1ε01ε01ε1ε01ε            8
                      1ε01ε01ε1ε01ε01ε1ε01ε1ε01ε01ε1ε01ε  13
             01ε1ε01ε1ε1ε01ε1ε01ε01ε1ε01ε01ε1ε01ε01ε1ε01ε1ε01ε  21
```

A further general property of L-grammars needs to be defined for internal relations between substrings within a grammar's derivation:

> A grammar G is *perfectly self-referential* iff any generation $g_x \in G$ can be expressed by means of *pure* concatenation relations between $g_i, g_j, \ldots, g_n \in G$ for *i, j, n* integers $< x$.

> A grammar G is *partially self-referential* iff any generation $g_x \in G$ can be expressed by means of a combination between concatenation relations and homomorphic mappings between $g_i, g_j, \ldots, g_n \in G$ for *i, j, n* integers $< x$.

These concepts are particularly significant for a topological approach to formal grammar, *as per* Mandelbrot's classification of fractals (also, Falconer, 2014), and we will see they also pertain to the concept of *structure preservation* in the sense of Emonds (1970), for phrase markers *qua* graphs resulting from the development of a rewriting rule system (as argued in Krivochen, 2017).

And finally, we will also define that a grammar $G = \{r_i, r_j, …, r_n\}$ over an alphabet $\Sigma = \{\sigma_1, \sigma_2, …\sigma_m\}$ is *exhaustive with respect to $\Sigma$* iff $i = j = n$ **and** there is at most a single occurrence of any $\sigma_x$ in both the left-hand side and the right-hand side of any *r*.

The notation used in the illustration of the Fib grammar above, which omits branches for graphical convenience, hides an important fact: the Fib grammar (and asymmetric L-grammars in general) growth pattern, which in this case is left-adjoining, can be exhaustively characterized in terms of a 'strong term' and a 'weak term', where the 'strong term' for any generation $g_n$ is that generated by the binary branching rule, and is a copy of $g_{n-1}$. The 'weak term' for $g_n$ is defined by the non-branching rule, and is a copy of $g_{n-2}$. As we said, such a grammar is *perfectly self-referential*. This means, crucially, that in infinite time, the weak term will 'catch up' with the main term, because *the residue contains the 'seed' of the main term*.

The permanence of the weak term in a lower dimensional space (Saddy, 2016), and the fact that the grammar is perfectly self-referential, justify the use of recurrence relations to formalize the grammar, such that:

17) $g_n = g_{n-2} \frown g_{n-1}$ (where the arc is interpreted, as usual, as linear concatenation)

This form of describing the $n^{th}$ Fib generation should not be surprising, for the $n^{th}$ Fib number is defined by the recurrence relation

18) $Fib_n = Fib_{n-2} + Fib_{n-1}$

This property of perfectly self-referential asymmetric L-systems, as we pointed out above, gives us a 'memory' for free, which is built into the system in the form of the weak term, which catches up to the strong term as $t \to \infty$.

4. A note on context-sensitivity

The relevance of contextual constraints for models of cellular automata, including L-systems, is well known (e.g., Farmer et al., 1984; also, notably, Conway's *Game of Life*). Mitchell et al. (1993) summarize environmental conditions for the development of cellular automata: given a binary alphabet $\Sigma = \{0, 1\}$ let $\eta$ be the set of possible neighbourhoods, let $\phi$ be a function that determines the behaviour of a target cell depending on the neighbourhood, and let $s = \phi(\eta)$ be the 'output bits', to which the central cell is updated:

| $\eta$ | 000 | 001 | 010 | 011 | 100 | 101 | 110 | 111 |
|---|---|---|---|---|---|---|---|---|
| $s$ | 0 | 0 | 0 | 1 | 0 | 1 | 1 | 1 |

Mitchell et al. (1993: 5) explain: *In words, this rule says that for each neighborhood of three adjacent cells, the new state is decided by a majority vote among the three cells*. Conway's *Game of Life* offers a similar perspective. Let '1' represent a living cell, and '0', a dead cell. The conditions are the following:

1. Any live cell with fewer than two live neighbours at *t* dies at *t+1*.
2. Any live cell with two or three live neighbours at *t* stays alive at *t+1*.
3. Any live cell with more than three live neighbours at *t* dies at *t+1*.
4. Any dead cell with exactly three live neighbours at *t* becomes alive at *t+1*.

We can see the similarity between Conway's conditions and the scenario presented in Mitchell et al. (1993): whether a given slot in the phase space for the cellular automaton is occupied by a living cell or a dead cell does not depend on properties of that slot alone and at a single moment in time, but both contextual conditions and the subsequent derivational step are crucial. In other words: the system is described by 'equations of motion' (Mitchell et al., 1993: 5), which have the form of recurrence relations. This is a clear difference with the Turing-computable picture:

> *The machine is supplied with a "tape" (the analogue of paper) running through it, and divided into sections (called "squares") each capable of bearing a "symbol". At any moment there is just one square, say the r-th, bearing the symbol $\mathfrak{S}(r)$ which is "in the machine". We may call this square the "scanned square". The symbol on the scanned square may be called the "scanned symbol". The "scanned symbol" is the only one of which the machine is, so to speak, "directly aware".* (Turing, 1936: 231)

Conditions over L-systems and their rules themselves make reference to two states: for a $g$ a generation, we need access to at least $g$ and $g-1$, by virtue of the nature of L-systems as models of biological growth. For a Turing machine to be able to access something else than the 'scanned symbol', the machine's $m$-configuration has to be affected, and a separate memory stack needs to be added (Turing, 1936). This is not the case for L-systems, which, because of their structure preserving properties (and the oscillatory engine they define), have a 'built-in' memory (Saddy, 2017).

5. <u>Further properties of L-languages</u>

In this work we have distinguished two mutually irreducible kinds of L-grammars: symmetric and asymmetric. Let an Asymmetric L-Language (ALL) be any language generated by an asymmetric L-grammar, and a Symmetric L-Language (SLL) be any language generated by a symmetric L-grammar. These languages can be characterized in terms of standard properties of formal languages (see Hopcroft and Ullman, 1969):

- Closed under union (?)
- Closed under intersection
- Closed under concatenation (?)
- Closed under Kleene star
- Closed under substitution (?)

We must take into consideration that these are properties which need to be proved for L-systems, and those whose applicability is relative to one or the other kind of L-grammar have been marked with (?), which only sets part of the future agenda for the present line of inquiry. These properties seem to hold *only within a kind of L-language*. They do not hold if we relate an ALL and a SLL. Thus, *the space of ALL and of SLL are disjoint*, and we can take this as a hard condition when parametrizing the L-space. It is worth pointing out that Union, Substitution, and Concatenation define an L-grammatical language, but not necessarily a Fib-grammatical language (one that conforms to the conditions on *n*-grams that emerge from Fib: *00, *111) if the input is the Fib grammar. Now, interestingly, Intersection and Kleene star closure hold for Fib grammatical strings, which impose stricter conditions than mere L-grammaticality. As is nearly a commonplace in this work, we need to make it explicit that these provisional claims are not analytical, and an analytical proof is currently being devised.

An important part of the parametrization of the space defined by L-systems, which is infinite but orthogonal to that defined by normal grammars, is defining a procedure to establish weak / strong equivalence between grammars. The 'tree pruning' operations formulated by Uriagereka and Krivochen above, and the converse expansion possibilities we have proposed here pertain specifically to this point. But graph-theoretic operations are only one way to approach this problem. Saddy has proposed in unpublished work that if the proportion of members of Σ between G and G' is the same, then G' and G are strongly equivalent in their quantitative properties at any generation $g_x$. That is,

19) 0 → 0 1
    1 → 1 0 1
20) 0 → 0 1 0 1
    1 → 1 0 1 1 0 1

At any generation, the ratio between the number of 0s and 1s in the grammars (19) and (20) is the same. This makes the grammars quantitatively equivalent.

There is a derivational side to this coin, which can roughly be formulated as follows: complex right-hand sides of L-rules can be either 'pruned' or expanded, by means of the mechanisms explained above. The idea is the following: if there exists a structure-preserving mapping between G and G' (that is, if there is a homomorphic mapping in either pruning or expansion terms), they are strongly equivalent. It must be noted that both the representational and the derivational approaches are not analytical, and further proof needs to be devised.

**Conclusion and further prospects:**

The main points of the present paper are simple and straightforward:

a) L-systems define a phase space which is orthogonal to that defined by Chomsky/Greibach-normal grammars, and which we have predictably called 'L-space'. These two are not mutually translatable; although Saddy (2017) has proposed that a dynamical system can go back and forth between both spaces. It is possible to 'normalize' an L-grammar under specific conditions of tree-pruning, as we saw in the proposals of Uriagereka and Krivochen, but then we have to give up aspects of the derivation of the L-grammar we are tampering with.

b) The L-space is *not* uniform or homogeneous: it can be further parametrized, by distinguishing mutually irreducible kinds of L-systems, which display different derivational and representational properties (see also Patel et al., 2015), as well as emergent behaviour. The topological properties of this space are currently being looked into, although it seems that, contrarily to the Chomsky-Normal space, it is non-metric (possibly ultrametric). The 'normalization' we have mentioned entails a metrization of the space as much as a modification of fundamental properties of the L-formalism (like the imposition of representational stability in the form of labels). Basically, there is a limited format for irreducible non-trivial L-grammars:

The rule rewriting the axiom can be:

i)   Axiom → non-axiom
ii)  Axiom → non-axiom, axiom
iii) Axiom → non-axiom, non-axiom

The rule rewriting the non-axiom can, in turn, be:

iv)  Non-axiom → axiom

v)   Non-axiom → axiom, non-axiom

vi)  Non-axiom → non-axiom, non-axiom

Not all combinations of these are useful, though. A grammar defined by (i)-(iv) just gives us an alternation of axiom/non-axiom in a unary branching tree. However, if the axiom rule branches as in (ii) (thus giving us a grammar like 0 → 0 1, 1 → 0 by (ii) and (iv)) we get strings that can be mapped via M / N to Fib strings:

0 → 0 1
1 → 0

$0 = Fib_0$
$10 = (Fib_2)^N = (Fib_2)^M$
$010 = (Fib_3)^N$
$10010 = (Fib_4)^N$
$00101001 = (Fib_5)^N$

(i) and (v) gives us the format of Fib

(ii) and (v) gives us the format of X-OR

(ii) and (vi) gives us an infinite string of non-axioms, so the grammar is not useful at all.

(iii) and (iv) gives us alternating strings of axiom and non-axiom symbols, again, not a very useful grammar.

(iii) and (v) gives us the format of Feigenbaum

And, finally, (iii) and (vi) gives us, again, alternating strings of axiom and non-axiom symbols.

c) The L-space presents a crucial difference with respect to the Chomsky-Normal space: only the latter displays subset relations between the formalisms it encompasses. Thus, for instance, a Turing Machine can generate any grammar within the Chomsky-Normal space, because such generative mechanisms properly contain CSG, CFG, and FSA. In the case of the L-space, the division between SLLs and ALLs seems to be primitive: neither is reducible to the other. The parametrization of the L-space we propose here defines two big areas, which are then in turn populated by grammars. Each of these areas (SL-grammars and AL-grammars) seem to constitute a *group* in the mathematical sense.

A strong interpretation of the parametrization of the L-space we have attempted here is that we effectively have exhaustively tessellated the L-space, meaning every L-system conceivable will fall into either of our categories: symmetric or asymmetric L-grammars. Moreover, any L-grammar which is more complex than the minimal expressions of either category will be reducible to the minimal symmetric or minimal asymmetric L-grammars by means of pruning, and conversely, the minimal exponents of each kind of grammar will be expandable into arbitrarily complex L-grammars. This, of course, is an empirical problem, which is currently being addressed on analytical basis.

**References:**

Brlek, Srećko (1989) Enumeration of factors in the Thue-Morse word. *Discrete Applied Mathematics*, 24. 83-96.

Chomsky, Noam (1957) *Syntactic Structures*. The Hague: Mouton.

(1959) On Certain Formal Properties of Grammars. *Information and Control* 2. 137-167.

Crutchfield, James and Karl Young (1989) Computation at the onset of chaos. In W. H. Zurek (ed.) *Complexity, Entropy, and the Physics of Information*. California: Addison-Wesley. 223-269. [available online at http://csc.ucdavis.edu/~cmg/papers/CompOnset.pdf]

Emonds, Joseph (1970) *Root and Structure Preserving Transformations*. PhD Thesis, MIT.

Falconer, Kenneth (2014) *Fractal Geometry*. London: Wiley.

Farmer, Doyne, Tomasso Toffoli and Stephen Wolfram (1984) (eds.) *Cellular Automata: Proceedings of an Interdisciplinary Workshop*. Amsterdam: North Holland.

Fortuny, Jordi (2009) *The emergence of order in syntax*. Amsterdam/Philadelphia: John Benjamins.

Geambasu, Andrea, Andrea Ravignani, and Clara C. Levelt (2016) Preliminary Experiments on Human Sensitivity to Rhythmic Structure in a Grammar with Recursive Self-Similarity. *Front. Neurosci.*, 28. http://dx.doi.org/10.3389/fnins.2016.00281

Greibach, Sheila (1965) A New Normal-Form Theorem for Context-Free Phrase Structure Grammars. *Journal of the ACM* **12** (1). 42-52.

Hopcroft, John and Jeffrey Ullman (1969) *Introduction to Automata Theory, Languages, and Computation*. Addison-Wesley.

Joshi, Aravind K. (1985) Tree adjoining grammars: How much context-sensitivity is required to provide reasonable structural descriptions? In David Dowty, Lauri Kartunnen, and Arnold Zwicky (eds.) *Natural Language Parsing*. Cambridge, Mass.: CUP. 206-250.

Joshi, Aravind K., and Tony Kroch (1985) Linguistic significance of TAG's. Ms.

Joshi, Aravind K., and Leon Levy (1982) Phrase structure trees bear more fruit than you would have thought. *American Journal of Computational Linguistic* 8(1). 1-11.

Krieger, Dalia (2007) On critical exponents in fixed points of non-erasing morphisms. *Theoretical Computer Science*, 376. 70–88

Krivochen, Diego (2017) *Aspects of Emergent Cyclicity in Language, Physics, and Computation*. PhD Thesis, University of Reading.

Lasnik, Howard and Joseph Kuppin (1977) A restrictive theory of transformational grammar. *Theoretical Linguistics* 4. 173–196.

Lees, Robert (1976) What are transformations? In James McCawley (Ed.) *Syntax and Semantics 7: Notes from the Linguistic Underground*. New York: Academic Press. 27-41.

Mandelbrot, Benoit (1983) *The fractal geometry of nature*. W. H. Freeman, San Francisco.

Matlach, Vladimir and Diego Krivochen (2016) Measuring string 'randomness': applications of quantitative methods to natural and formal languages. Presented at OLINCO, June 9[th], 2016.

Mitchell, Melanie, James Crutchfield, and Peter Hraber (1993) Dynamics, Computation, and the "Edge of Chaos": A Re-Examination. In G. Cowan, D. Pines, and D. Melzner (eds.) *Integrative Themes*. Reading, MA. Addison-Wesley.

Patel, Ebrahim, Peter Grindrod, Douglas Saddy, and Andrew Irving (2015) L-systems: Classification and Equivalence. Ms. University of Oxford, University of Reading, and University of Manchester.

Phillips, Bethany (2017) *Symmetry of Hierarchical Artificial Grammars and Its Effects on Implicit Learning*. MSc Thesis: University of Reading.

Prusinkiewicz, Przemyslaw and Aristid Lindenmayer (2010) *The Algorithmic Beauty of Plants*. Springer-Verlag. [2nd edition]

Prusinkiewicz, Przemyslaw and James Hanan (1989) *Lindenmayer Systems, Fractals, and Plants*. Springer-Verlag.

Ross, John Robert (1967) *Constraints on Variables in Syntax*. PhD Thesis, MIT.

(1969) A proposed rule of Tree-Pruning. En David Reibel & Sanford Schane (eds.) *Modern Studies in English: Readings in Transformational Grammar*. New Jersey: Prentice Hall. 288-299.

Saddy, Douglas (2009) Perceiving and Processing Recursion in Formal Grammars. Recursion: Structural Complexity in Language and Cognition Conference. UMass.: Amherst.

(2017) Syntax and Uncertainty. To appear in Roger Martin (ed.) *Language, Syntax, and the Natural Sciences*. Cambridge: CUP.

Schmerling, Susan F. (1983) Two theories of syntactic categories. *Linguistics and Philosophy* 6. 393–421.

Schmid, Samuel (2017) Bilingualism and Grammar Induction. Presentation, University of Geneva.

Thue, Axel (1912) Uber die gegenseitige Lage gleicher Teile gewisser Zeichenreihen, *Norske vid. Selsk. Skr. Mat. Nat. Kl.*, 1. 1–67.

Turing, Alan (1936) On computable numbers, with an application to the Entscheidungsproblem. *Proceedings of the London Mathematical Society* 42 (2), 230-265.

Uriagereka, Juan (2015) "Natural" grammars and natural language. In Beatriz Fernandez and Pello Salaburu (eds.) *Ibon Sarasola, Gorazarre*. Bilbao: Editorial de la Universidad del País Vasco. 665-674.

Vitányi, Paul M. B. (1978) *Lindenmayer Systems: Structure, Languages, and Growth Functions*. Amsterdam: Matematisch Centrum.